\documentclass[aps,prapplied,12pt,superscriptaddress,notitlepage]{revtex4-2} 
\usepackage[utf8]{inputenc}
\usepackage{amsmath, amssymb, bm}      % Math packages
\usepackage{graphicx}                   % Graphics
\usepackage{wrapfig}                    % Wrapping text around figures
\usepackage{array}                       % Array and tabular enhancements
\usepackage{booktabs}                   % Better tables
\usepackage{setspace}                   % Line spacing
\usepackage{placeins}                   % Float control
\usepackage{chngcntr}                   %change fig cntr for Supp.
\usepackage{braket}

\UseRawInputEncoding
\usepackage{xr}                         % External references

\usepackage{tikz}                       % TikZ package
\usetikzlibrary{trees}                  % TikZ libraries
\begin{document}

\title{Quasi-bandgap behavior in non-Hermitian photonic crystals}

\author{Jin Xu}\affiliation{Department of Materials Science and Engineering, University of California, Los Angeles, Los Angeles, CA 90095, USA}

\author{Daniel Cui}\affiliation{Department of Materials Science and Engineering, University of California, Los Angeles, Los Angeles, CA 90095, USA}
\author{Aaswath P. Raman}
\email{aaswath@ucla.edu}
\affiliation{Department of Materials Science and Engineering, University of California, Los Angeles, Los Angeles, CA 90095, USA}

\begin{abstract}
We investigate non-Hermitian photonic crystals in which the lossy and lossless constituents share the same real permittivity and differ only in their imaginary part. We characterize the complex band structure and reflection response of both one-dimensional (1D) and two-dimensional (2D) systems, and show that introducing even a small amount of material loss opens a quasi bandgap at the Brillouin-zone boundary. This quasi bandgap, absent in the lossless limit of the same structure, gives rise to sharp reflectivity peaks whose origin we explain through second-order perturbation theory. As an application of this behavior, we demonstrate a selective reflector combining a conventional photonic-crystal waveguide with a non-Hermitian photonic crystal, achieving wavelength-selective reflection with broadband absorption.
\end{abstract}
\maketitle

\section{Introduction}
The emergence of photonic bandgaps in optical nanostructures has motivated decades of fundamental and applied research in nanophotonics. At the heart of this phenomenon was the observation that wavelength-scale periodic nanostructures that exhibited a sufficiently strong contrast in the real part of their dielectric permittivity exhibit bandgaps for photons analogous to those seen by electrons in electronic systems. These bandgaps in turn have enabled a rich array of physical mechanisms, including slow light, band-edge density of state enhancement, defect states and more \cite{joannopoulos2008molding,raman2010photonic,fan2002analysis,karabchevsky2020chip}. Due to these capabilities they have become an indispensable technology across the entire field of optical physics and photonics \cite{zhang2018review,dutta2016coupling,pinto2012photonic,soukoulis2012photonic}.

Photonic crystals are typically assumed to be in Hermitian in nature \cite{joannopoulos2008molding} due to their constituent materials being effectively lossless over target wavelengths of operation. This is desirable for many conventional applications and photonic devices. Over the last decade however there has been significant interest in studying non-Hermitian photonic systems for applications in lasers and signal processing\cite{takata2021observing,hodaei2014parity}. Hermitian operators formally can only describe idealized, lossless systems. However, non-conservative elements (e.g., radiation losses, material absorption) are ubiquitous in real-world system. For example, when light propagates in media with a complex refractive index, the system is inherently non-Hermitian\cite{feng2017non, longhi2018parity}. Perhaps the most studied non-Hermitian systems are parity-time ($\mathcal{PT}$) symmetric systems\cite{bender1998real,chong2011p,longhi2010pt,ruter2010observation,alaeian2014non,zhen2015spawning,cerjan2016exceptional,benisty2011implementation, chen2017exceptional,merkel2018unidirectional,ozdemir2019parity,zhou2019exceptional,miri2019exceptional,cerjan2019experimental,bender2002complex,el2007theory}.
A $\mathcal{PT}$ symmetric optical potential requires a balanced distribution of gain and loss in the medium. Experimentally realized $\mathcal{PT}$-symmetric systems have primarily involved direct coupling of optical resonance in waveguide \cite{feng2014single, zhao2016metawaveguide,xu2016experimental,lawrence2014manifestation, peng2014parity,guo2009observation,ruter2010observation,regensburger2012parity,wen2019experimental} or deliberately designed 2D systems \cite{kremer2019demonstration,biesenthal2019experimental,takata2021observing}.  

In the context of photonic crystals, the role of non-Hermiticity has also had substantial exploration. Early studies about polaritonic photonic crystals or lossy photonic crystals\cite{huang2004nature, davancco2007complex,naito2008experimental} involve systems where the two constituent materials have substantially different real permittivities, so that the observed bandgaps originate primarily from real index contrast rather than from loss alone. However these past works examined systems with large real-permittivity contrast supplemented by dispersive loss. By contrast, recent work on non-Hermitian photonic crystals have explicitly explored the band structure consequences of engineering $\mathcal{PT}$-symmetry through the imaginary permittivity distribution of the photonic crystal \cite{cerjan2016exceptional, cerjan2016effects, zhou2019exceptional}. In this context, the emergence of quasi-bandgaps in these systems has been observed \cite{cerjan2016effects}, indicating that loss alone can play a role in their opening and behavior. However, a systematic understanding of the nature and origin of the quasi-bandgaps driven by the imaginary permittivity remains lacking.

Here we investigate non-Hermitian photonic crystal systems in which the constituent materials share the same real permittivity but differ in their imaginary permittivity and characterize their complex band structure and reflection response. We find that a quasi bandgap emerges at the edge of the Brillouin zone, characterized by a sharp reflection peak that is observed in both 1D and 2D non-Hermitian photonic crystals. We develop a second-order perturbation theory the elucidates the origin of the quasi-bandgap and can predict the bandgap width, showing that it scales quadratically with the imaginary permittivity introduced into one of the two constituent materials. As an application of this framework, we design a selective reflector which is made of a lossless 2D photonic crystal waveguide and a non-Hermitian photonic crystal that reflects light at the quasi bandgap while otherwise absorbing light at other wavelengths. Our results highlight the unique behavior that loss alone can yield in periodic photonic systems.

\section{Quasi-Bandgaps in 1D Non-Hermitian Photonic Crystals}
\begin{figure}[h!]
    \centering
    \includegraphics[width = 16cm]{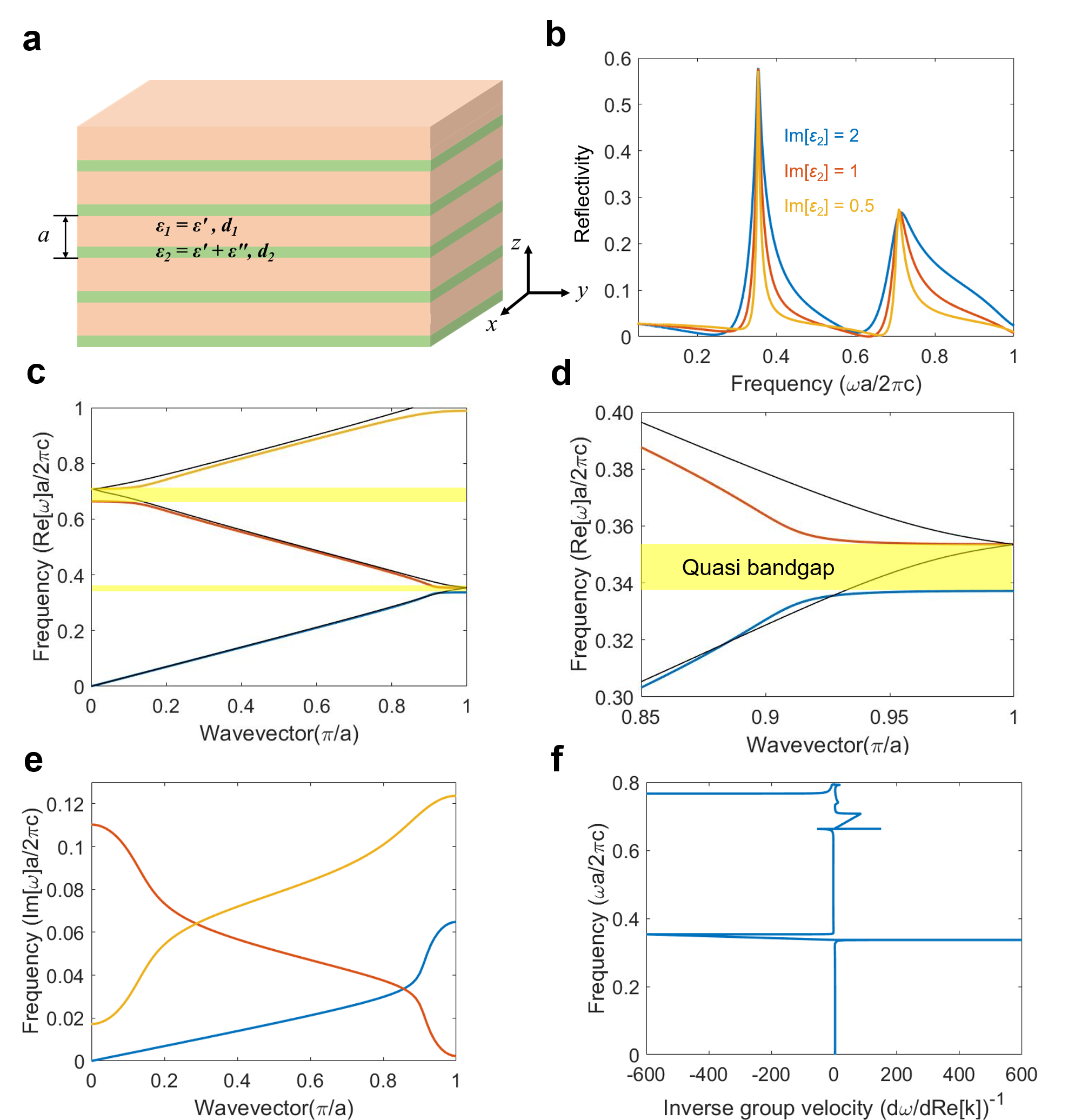}
    \caption{1D non-Hermitian photonic crystal: (a) schematic diagram, (b) reflectivity for three values of $\mathrm{Im}[\varepsilon_2]$. A reflectivity peak of 0.57 is observed at a scaled frequency 0.35 $(2\pi c/a)$, (c) real part of the band structure computed assuming a real wavevector (colored lines) and a real frequency (thin black lines). (d) enlarged view of the first and second bands showing the quasi bandgap around frequency 0.35, (e) imaginary part of the band structure assuming real wavevector, (f) density of state / inverse group velocity $(d\omega/d \mathrm{Re}[k])^{-1}$ for the bands and frequency range examined.}
    \label{fig:Reflection of the 1D non-Hermitian Photonic Crystal system}
\end{figure}

We first consider the behavior of a canonical 1D non-Hermitian photonic crystal shown in Fig.~\ref{fig:Reflection of the 1D non-Hermitian Photonic Crystal system}(a). The photonic crystal consists of alternating layers of materials (orange and green) whose real permittivities are identical but imaginary permittivities vary ($\varepsilon_1 = 2, \varepsilon_2 = 2 + 2i$), with a spatial period of $a$ and $d_1 = 0.8a$, $d_2 = 0.2a$. Each layer is uniform and extends to infinity along the $x$ and $y$ directions. Fig.~\ref{fig:Reflection of the 1D non-Hermitian Photonic Crystal system}(b) shows the reflectivity of a 50 bilayer 1D non-Hermitian photonic crystal when $\varepsilon_1 = 2$, and $\varepsilon_2 = 2 + 0.5i$, $2 + 1i$ and $2 + 2i$ respectively. Though only the imaginary part of the permittivity is different, we observe narrow reflectivity peaks with an amplitude of 0.57 at a frequency of 0.35 $(2\pi c/a)$, and an amplitude of 0.27 at 0.71 $(2\pi c/a)$ for all three values of $\mathrm{Im}[\varepsilon_2]$. A larger $\mathrm{Im}[\varepsilon_2]$ increases the bandwidth of the reflectance peak. 

These results share some similarities with the bandgaps observed in conventional Hermitian photonic crystals, however with important differences as well. To elucidate their behavior we compute the band structure of this 1D non-Hermitian photonic crystal by solving the standard transcendental equation for both the wavevector $k$ given a frequency $\omega$, and for $\omega$ when $k$ is given. If the frequency is assumed to be real, the resulting wavevector is complex since the permittivities are complex and the wave decays in space, while if the wavevector is assumed to be real the resulting frequency is complex and the real wavevector excitation decays in time. The two solutions are only identical when $\varepsilon$ is real.

We plot the real component of the solutions assuming a real $\omega$ or a real $k$ in Fig.~\ref{fig:Reflection of the 1D non-Hermitian Photonic Crystal system}(c). The colored curves are the solution for real wavevector while the black thinner lines are the solutions for the real frequency case. We can see that near the band edges, the two methods produce different results. When $k$ is real, we clearly observe two bandgaps where no solution exists within these frequency ranges: one is around a scaled frequency of 0.35 and the other around a scaled frequency of 0.71 which corresponds to the two reflectivity peaks we see in Fig.~\ref{fig:Reflection of the 1D non-Hermitian Photonic Crystal system}(b). However, the band structure solved by assuming real frequency is a continuous line, with a solution lying somewhere within the Brillouin zone for all frequencies. This is a signature of a \emph{quasi} bandgap that has previously been observed in lossy polaritonic photonic crystals \cite{huang2004nature}.

Fig.~\ref{fig:Reflection of the 1D non-Hermitian Photonic Crystal system}(d) shows the enlarged first and second bands. Here we can see more clearly that there is a quasi bandgap between the first and second band as solved by real wavevector. The imaginary part of the first three bands solved by assuming real wavevector exhibits minima and maxima at the band edge (in Fig.~\ref{fig:Reflection of the 1D non-Hermitian Photonic Crystal system}(e)). This implies that the wave experiences either high loss or low loss at different band indices. To show the slow light at the band edges, the inverse group velocity $(d\omega/d\mathrm{Re}[k])^{-1}$ obtained from the real frequency solutions is shown in Fig.~\ref{fig:Reflection of the 1D non-Hermitian Photonic Crystal system}(f). We can see the inverse group velocity increases significantly near the quasi bandgap region.

After identifying the quasi bandgaps purely induced by loss in the 1D non-Hermitian photonic crystal, we seek to explain the emergence of these quasi bandgaps by developing and extending the conventional perturbation theory used for lossless photonic systems. First order perturbation theory is widely used in lossless perturbed photonic crystal systems to predict the imaginary eigenvalue change from introducing loss to the unperturbed system \cite{johnson2002perturbation}. Higher order perturbation theory has, to our knowledge, not been used to examine the effect of loss on the real part of a mode's eigenfrequency. Here, we employ both first and second order perturbation theory for the generalized electromagnetic wave eigenvalue problem to predict both the imaginary and real bandgap size for the 1D non-Hermitian photonic crystal described in Fig.~1. We note that the unperturbed system ($\Delta\varepsilon = 0$) is a homogeneous dielectric with $\varepsilon = 2$, whose band structure is the folded free-photon dispersion with no bandgap. The quasi bandgap therefore arises entirely from the imaginary-permittivity perturbation.

\section{Second-Order Perturbation Theory Correction}
We begin with the electric field formulation of Maxwell's equations formulated as an eigenproblem, where the electric field with time dependence $e^{-i\omega t}$ is in a source-free linear dielectric medium $\varepsilon(r)$:

\begin{equation}
    \nabla \times \nabla \times E(r) = \left( \frac{\omega}{c} \right)^2 \varepsilon(r) E(r) \label{eq:EMwave}
\end{equation}

Assuming that $\varepsilon(r)$ is purely real and lossless and positive, the eigenproblem is Hermitian and positive semidefinite, leading to real $\omega$ solutions. Since it is a generalized eigenproblem, the eigenstates are orthogonal under the inner product. We then consider a small parameter $\delta$ characterizing the perturbation, which in our case will be $\Delta \varepsilon$: a change in the imaginary part of the permittivity. The new eigensolutions are expanded in powers of $\Delta \varepsilon$. The first-order correction from a perturbation $\Delta\varepsilon$ is then found to be the standard result

\begin{equation}
    \Delta \omega^{(1)} = - \frac{\omega^{(0)}}{2} \frac{\braket{\mathbf{E}_n^{(0)}| \Delta \epsilon | \mathbf{E}_n^{(0)}}}{\braket{\mathbf{E}_n^{(0)}| \epsilon_0 | \mathbf{E}_n^{(0)}}}
    \label{eq:1st_order}
\end{equation}

where $\Delta\omega^{(1)}$ is the first order correction for the perturbed eigenvalue, $\omega^{(0)}$ and $\mathbf{E}_n^{(0)}$ are the unperturbed eigenvalue (eigenfrequency) and eigensolution (electric field). Since in the non-Hermitian photonic crystals we consider the perturbation is purely imaginary the frequency shift is purely in the imaginary frequency as well. However, as observed in Fig. 1, a real frequency shift associated with the quasi-bandgap is present as well. \\
To find the real frequency correction, a $|\Delta\varepsilon|^{2}$ term must be present. We expect this term to be present in the second order correction, which we develop (see Supplementary Information) and find to be:

\begin{align}
     \Delta \omega^{(2)} = &\frac{3\omega^{(0)}}{8} \left( \frac{\braket{\mathbf{E}_n^{(0)}|\Delta \epsilon|\mathbf{E}_n^{(0)}}}{\braket{\mathbf{E}_n^{(0)}| \epsilon_0 |\mathbf{E}_n^{(0)}}} \right)^2 + \frac{\omega^{(0)}}{2} \sum_{\substack{j=1 \\ j\neq n}}^{N} \left[ \left( \frac{\lambda_n^{(0)}}{\lambda_n^{(0)} - \lambda_j^{(0)}} \right) \frac{\braket{\mathbf{E}_j^{(0)}| \Delta \epsilon |\mathbf{E}_n^{(0)}}}{\braket{\mathbf{E}_j^{(0)}| \epsilon_0 |\mathbf{E}_j^{(0)}}} \frac{\braket{\mathbf{E}_n^{(0)}| \Delta \epsilon |\mathbf{E}_j^{(0)}}}{\braket{\mathbf{E}_n^{(0)}| \epsilon_0 |\mathbf{E}_n^{(0)}}} \right]
     \label{eq:2nd_order}
\end{align}
Eq.~\eqref{eq:2nd_order} is the primary result of our theoretical analysis. We note that this result immediately satisfies the requirement that a purely imaginary frequency change in the permittivity should yield a real frequency shift. Furthermore, it highlights a quadratic dependence on the imaginary permittivity value in the real frequency shift one would observe.

To provide further physical intuition, we consider as a limiting case the single-mode approximation and assume standing wave electric field profiles of $\mathbf{E}^{(0)} = \cos(\pi x/a)$ and $\mathbf{E}^{(0)} = \sin(\pi x/a)$ for any two sequential bands. These standing-wave profiles correspond to the degenerate Bloch modes at the zone boundary ($k = \pi/a$) of the unperturbed homogeneous medium, whose degeneracy is lifted by the imaginary-permittivity perturbation. We can then calculate the size of the real frequency quasi-bandgap size using Eq.~\ref{eq:2nd_order} (see Supplementary Information) to find:

\begin{equation}
    \Delta\omega^{(2)}_{gap} = \frac{3\omega^{(0)}}{2\pi} \left | \frac{\Delta \epsilon}{\epsilon_0}\right|^2 \frac{d}{a} \sin\left(\frac{\pi d}{a}\right)
\end{equation}

\begin{figure}[h!]
    \centering
    \includegraphics[width = 6in]{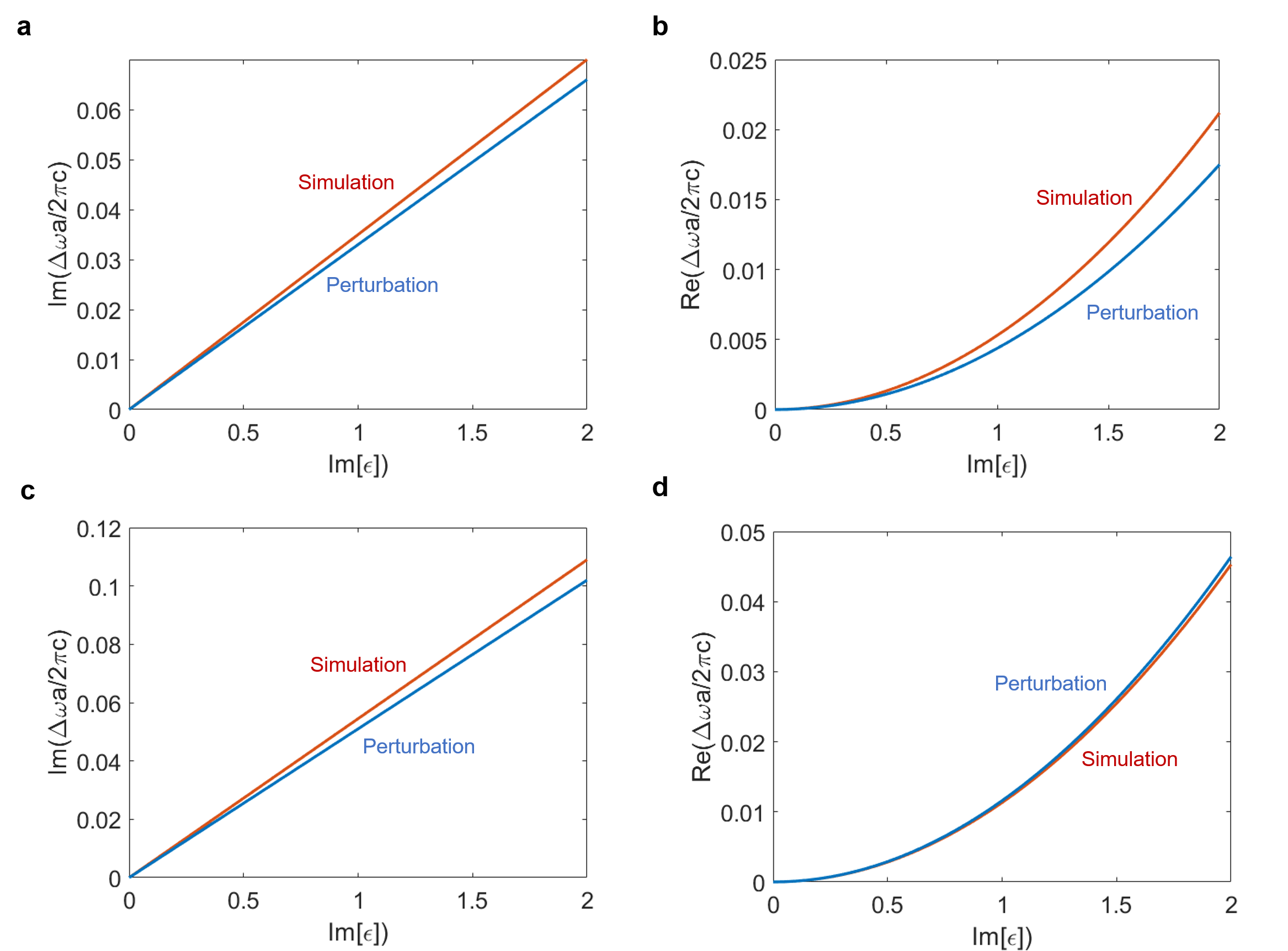}
    \caption{Perturbation theory prediction for the quasi bandgap size. (a) Imaginary bandgap size between the first and second band from $1^{st}$ order perturbation theory; (b) real bandgap size between the first and second band from $2^{nd}$ order perturbation theory; (c) imaginary bandgap size between the second and third band from $1^{st}$ order perturbation theory; (d) real bandgap size between the second and third band from $2^{nd}$ order perturbation theory.}
    \label{fig:perturbation}
\end{figure}
We note here again the quadratic dependence in the change in the permittivity, which in the case of a purely imaginary perturbation results in a real mode frequency shift. We plot the bandgap size between the first and second band and the second and third band in Fig.~\ref{fig:perturbation}. Fig.~\ref{fig:perturbation}(a) shows the imaginary bandgap size between the first and second band and (c) between the second and third bands from $1^{st}$ order perturbation theory. Both the perturbation theory and transfer matrix simulation results show that the imaginary bandgap size grows linearly with the perturbation $\Delta\varepsilon = \mathrm{Im}[\varepsilon]$. Fig.~\ref{fig:perturbation}(b) shows the real bandgap size between the first and second band and Fig.~\ref{fig:perturbation}(d) between the second and third band from $2^{nd}$ order perturbation theory. We see that the real bandgap opens when we introduce the perturbation (loss in the photonic crystal) and it grows quadratically with $\mathrm{Im}[\varepsilon]$. The real bandgap size becomes larger as $\mathrm{Im}[\varepsilon]$ increases, which explains why the reflection peak broadens at larger $\mathrm{Im}[\varepsilon]$ in Fig.~\ref{fig:Reflection of the 1D non-Hermitian Photonic Crystal system}(b). In addition, in Fig.~\ref{fig:perturbation}, all perturbation theory predictions of the bandgap size agree well with the simulation results and begin to deviate as $\mathrm{Im}[\varepsilon]$ becomes larger.

\section{2D Non-Hermitian Photonic Crystal Behavior}

\begin{figure}[h!]
    \centering
    \includegraphics[width = 14cm]{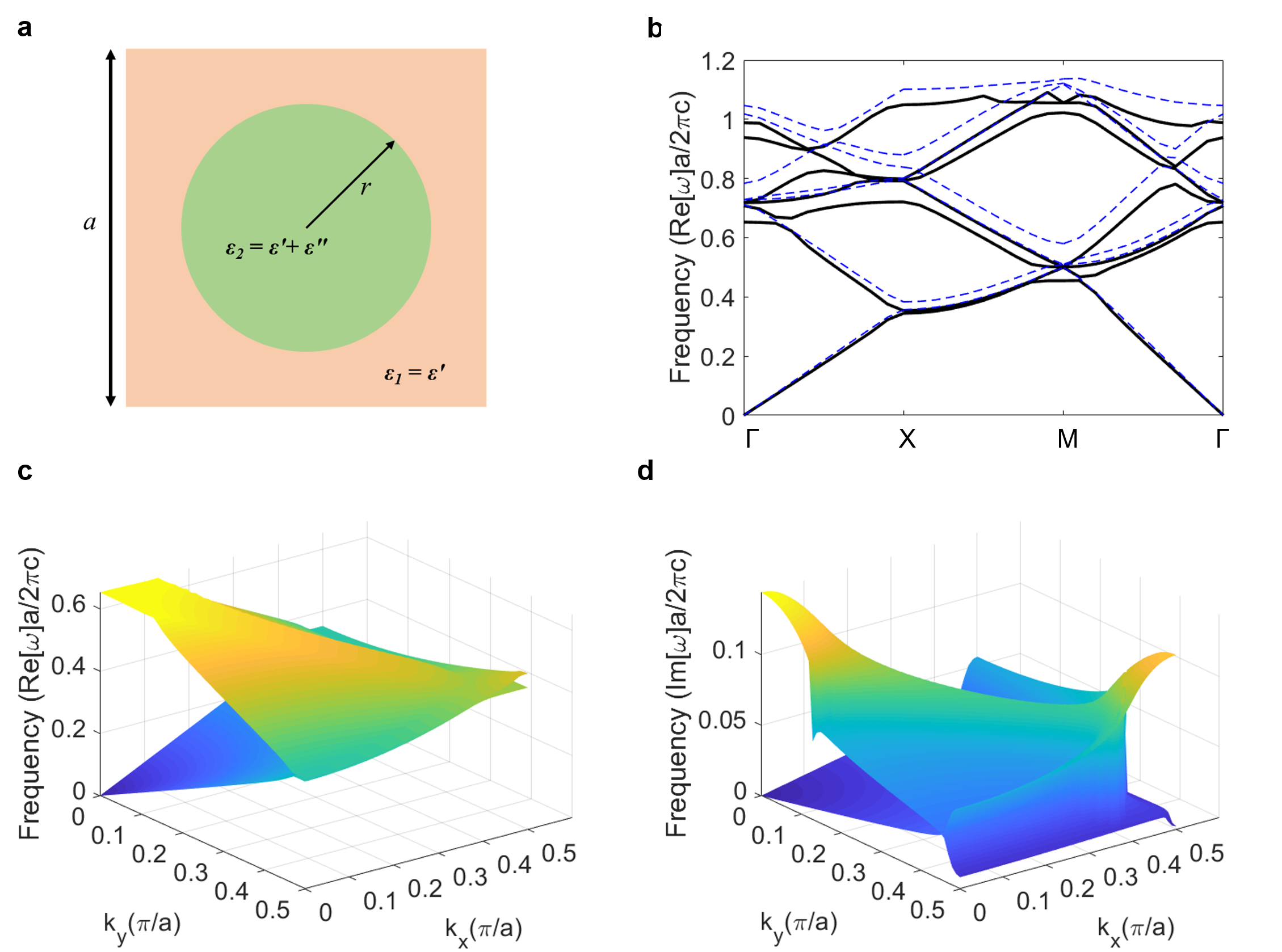}
    \caption{2D non-Hermitian photonic crystal. (a) Schematic diagram of the 2D non-Hermitian photonic crystal with square lattice lossy rods ($r = 0.21a$, where $a$ is the lattice constant) with $\varepsilon_2 = 2 + 2i$ in a lossless background $\varepsilon_1 = 2$; (b) TM band structure from $\Gamma$-$X$-$M$-$\Gamma$ (dashed blue lines show the band structure for the corresponding Hermitian system: homogeneous permittivity of 2); (c) real part of the band structure at all combinations of $k_x$ and $k_y$; (d) imaginary part of the band structure at all combinations of $k_x$ and $k_y$.}
    \label{fig:2Dband}
\end{figure}

We apply the insights we obtain from the 1D non-Hermitian photonic crystal to an analogous two-dimensional system. We consider a 2D non-Hermitian photonic crystal of lossy rods ($r = 0.21a$) with $\varepsilon_2 = 2 + 2i$ embedded in a lossless background $\varepsilon_1 = 2$ in a square lattice (lattice constant $a$) as shown in Fig.~\ref{fig:2Dband}(a). In Fig.~\ref{fig:2Dband}(b), we plot the TM band structure from $\Gamma$-$X$-$M$-$\Gamma$. We overlay (dashed blue lines) the band structure for the corresponding Hermitian or lossless photonic crystal (essentially a bulk material with homogeneous permittivity $\varepsilon = 2$). Compared with the band structure of the lossless crystal, the bands of the non-Hermitian photonic crystal are more complex where we see bands tend to split and merge more frequently at the high-symmetry points in the first reduced Brillouin zone. To better characterize the band structure features of the non-Hermitian photonic crystal, we plot the first two bands over a range of $k_x$ and $k_y$ in Fig.~\ref{fig:2Dband}(c) and (d). As we can see from Fig.~\ref{fig:2Dband}(c), the first band and second band tend to merge at the band edge but there is still a gap between them. On the other hand, different bands may experience low loss or high loss at the same $k$ points, as shown in Fig.~\ref{fig:2Dband}(d).

\begin{figure}[h!]
    \centering
    \includegraphics[width = 10cm]{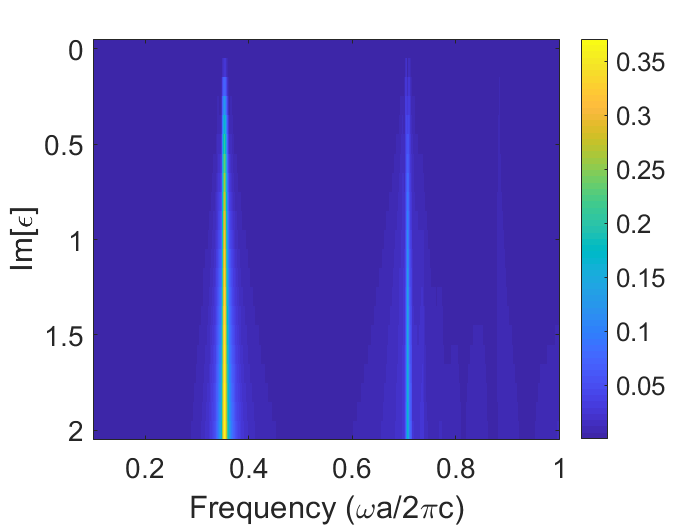}
    \caption{Reflectivity of the 2D non-Hermitian photonic crystal for TM mode at normal incidence. The reflectivity has a sharp peak at frequency 0.35 and a weaker peak at 0.71, analogous to the 1D case.}
    \label{fig:2Dreflectivity}
\end{figure}

In Fig.~\ref{fig:2Dreflectivity}, we calculate the reflectivity of the 2D non-Hermitian photonic crystal for TM mode. The reflectivity has a remarkably sharp peak at frequency 0.35. The reflectivity is purely induced by the loss in the pillars and increases with the growth of the imaginary permittivity $\mathrm{Im}[\varepsilon_2]$ of the lossy pillars and stabilizes when $\mathrm{Im}[\varepsilon_2]$ is greater than 0.5, which is similar to the behavior observed in the analogous 1D photonic crystal. This peak is also located within the gap at the band edge between the first and the second band as we observed in Fig.~\ref{fig:2Dband}(c). There is also a weaker peak around a frequency of 0.71.

\begin{figure}[h!]
    \centering
    \includegraphics[width = 16cm]{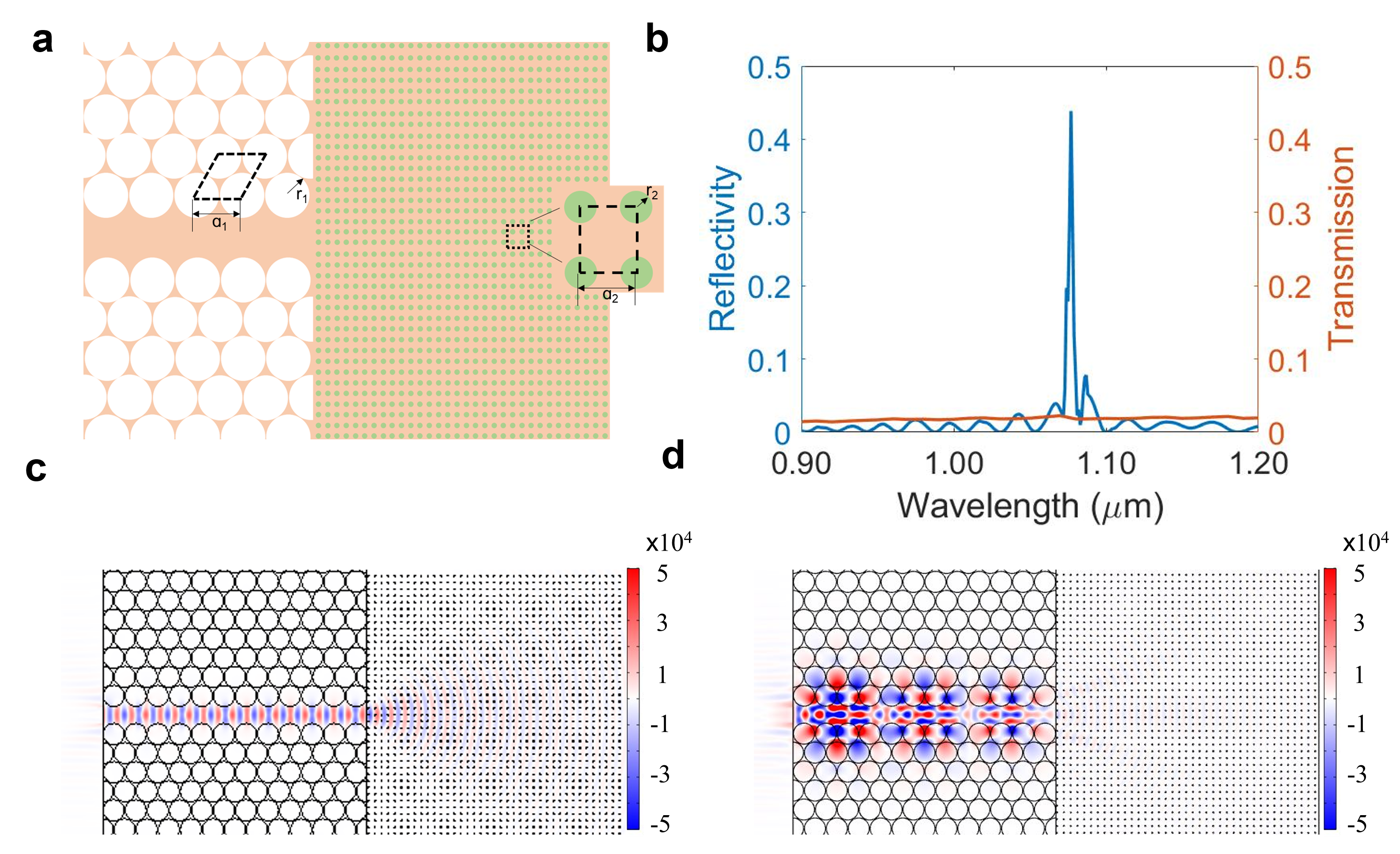}
    \caption{A selectively reflective waveguide. (a) Schematic diagram of the reflector: the left part is a 2D triangular air hole photonic crystal with a linear defect; the right part is a non-Hermitian square lattice photonic crystal with a square array of lossy pillars ($\varepsilon = 13 +6i$) embedded in a lossless substrate with the same real part of permittivity ($\varepsilon = 13$). (b) Reflectivity and transmission of the combined selective reflector from $0.9\,\mu$m to $1.2\,\mu$m. (c) and (d) Out-of-plane electric field $E_z$ distribution at wavelengths $1.13\,\mu$m and $1.077\,\mu$m, respectively.}
    \label{fig:NHPC}
\end{figure}

Motivated by this behavior, and as an illustration of the potential utility of these phenomena, we design a selectively reflective photonic crystal waveguide architecture (Fig.~\ref{fig:NHPC}(a)), which can allow a broad wavelength range of light to pass through the left part of the geometry and be absorbed in the right half while only light at a designed wavelength will be reflected back to the left. The left part of the selective reflector is a 2D photonic crystal with a triangular array of air holes in a dielectric substrate ($\varepsilon = 13$) with a relative radius $r_1/a_1$ of 0.48, where $a_1$ is the lattice constant of the photonic crystal. This photonic crystal has a linear defect (one row of air holes is removed from the crystal). The 2D photonic crystal without the linear defect has a complete band gap with a midgap ratio of 16.3\%, which allows this waveguide to guide a wide range of light along the linear defect.

The right part of this reflector is a non-Hermitian square lattice photonic crystal with a square array of lossy pillars ($\varepsilon = 13 + 6i$) embedded in a lossless substrate with the same real part of permittivity ($\varepsilon = 13$). The relative radius $r_2/a_2$ is 0.1, where $a_2$ is the lattice constant of the non-Hermitian photonic crystal. The non-Hermitian photonic crystal on the right is carefully designed so that it has a reflection peak (similar to what we observed in Fig.~\ref{fig:2Dreflectivity}) within the guided wavelength range of the waveguide on the left.

As shown in Fig.~\ref{fig:NHPC}(b), the reflectivity of the combined selective reflector is ultra-low ($\ll 0.1$) at all wavelengths that pass through the waveguide on the left, except at the wavelength $1.077\,\mu$m. The reflectivity reaches 0.43 at $1.077\,\mu$m. A 20-fold contrast is achieved compared with the reflectivity of light far from $1.077\,\mu$m. The transmission throughout the wavelength range from $0.9\,\mu$m to $1.2\,\mu$m is below 0.025. With absorption $= 1 -$ reflectivity $-$ transmission, the majority of the light is absorbed if not reflected by the non-Hermitian photonic crystal part, making it distinct from a Bragg mirror, which reflects selectively but transmits the remaining light rather than absorbing it.

We show the out-of-plane electric field $E_z$ distribution in Fig.~\ref{fig:NHPC}(c) and (d) at two different wavelengths ($1.13\,\mu$m and $1.077\,\mu$m respectively). At $1.13\,\mu$m (Fig.~\ref{fig:NHPC}(c)), the light is guided along the 2D photonic crystal waveguide on the left and then penetrates into the non-Hermitian photonic crystal region where it experiences strong absorption within the first few columns of lossy pillars embedded in the substrate. In contrast, in Fig.~\ref{fig:NHPC}(d) at the reflection peak wavelength $1.077\,\mu$m, we can see that after the light passes through the left waveguide, it is reflected immediately as it enters the non-Hermitian region and barely penetrates it. This is exactly the behavior expected within a photonic bandgap, where light is blocked and cannot propagate. This observation further supports the presence of the quasi bandgap purely induced by loss in the non-Hermitian photonic crystal. This type of device may find applications in imaging and sensing \cite{threm2012photonic,bock2007optical,nair2010photonic,zhang2015review,zhuo2015label,pitruzzello2018photonic}, where broadband optical suppression by absorption with selective reflection may enable improved signal-to-noise performance.

\section{Conclusion}
 
The central finding of this work is that a periodic modulation of material loss alone, without any real index contrast, is sufficient to open quasi bandgaps in photonic crystals. Our second-order perturbation theory reveals the mechanism: while the imaginary permittivity perturbation shifts each mode's frequency along the imaginary axis at first order, it produces a real frequency splitting at second order that scales as $|\Delta\varepsilon|^2$. This quadratic scaling, confirmed by transfer matrix and full-wave simulations across both 1D and 2D geometries, explains both the existence and the broadening of the sharp reflectivity peaks observed within the quasi bandgap. The selective reflector we demonstrated achieves wavelength-selective reflection with broadband absorption, and illustrates how this loss-driven bandgap mechanism enables device functionality distinct from conventional Bragg mirrors.
 
Our results highlight the unique and counterintuitive capabilities possible by engineering loss in photonic architectures. This perspective opens several directions for future work, including the extension of loss-driven bandgap design to three-dimensional architectures, the exploration of disorder and topological effects in non-Hermitian photonic crystals, and the development of devices that exploit the interplay between quasi bandgaps and gain media. Additionally, the selective absorption-reflection capability demonstrated here may find immediate applications in sensing and imaging, where suppressing background light through absorption rather than scattering can improve signal-to-noise performance. 

\section{Acknowledgments}
This material is based upon work supported by the National Science Foundation (NSF CAREER) under Grant No. 2146577, and the Sloan Research Fellowship (Alfred P. Sloan Foundation).
\bibliography{Ref}

\end{document}